\documentclass[12pt]{iopart}
\usepackage{color}
\usepackage{graphics}
\usepackage{epsfig}
\usepackage{graphicx}
\usepackage{amssymb}
\usepackage[latin1]{inputenc}
\usepackage[T1]{fontenc}
\usepackage{color}

\newcommand{\slsh}[1]{{\not \! #1}}

\newcommand{\bea}{\begin{eqnarray}}
\newcommand{\eea}{\end{eqnarray}}

\begin{document}
\title[Impact of a magnetic field and temperature on explicit $\chi$SB in QED]{Impact of a 
uniform magnetic field and nonzero temperature on explicit chiral symmetry breaking in QED: Arbitrary hierarchy of energy scales}  

\author{A Ayala$^{1,2,3}$, A Bashir$^1$, A Raya$^1$ and A S\'anchez$^1$}   
\address{$^1$Instituto de F\1sica y Matem\'aticas,
Universidad Michoacana de San Nicol\'as de Hidalgo, Apartado Postal
2-82, Morelia, Michoac\'an 58040, M\'exico.}
\address{$^2$Instituto de Ciencias Nucleares, Universidad
Nacional Aut\'onoma de M\'exico, Apartado Postal 70-543, M\'exico
Distrito Federal 04510, M\'exico.}
\address{$^3$Centro Brasileiro de Pesquisas Fisicas, CBPF-DPC,
Rua Dr. Xavier Sigaud 150-Urca, 22290-180,
Rio de Janeiro, Brazil.}
\eads{\mailto{ayala@nucleares.unam.mx}, \mailto{adnan@ifm.umich.mx}, \mailto{raya@ifm.umich.mx}, \mailto{ansac@ifm.umich.mx}}

\begin{abstract} 
Employing the Schwinger's proper-time method, we calculate the $\langle\bar{\psi} 
\psi\rangle$-condensate for massive Dirac fermions of charge $e$ interacting with a uniform 
magnetic field in a heat bath.  We present general results for arbitrary hierarchy of the 
energy scales involved, namely, the fermion mass $m$, the magnetic field strength $\sqrt{eB}$ 
and temperature $T$. Moreover, we
study particular regimes in detail and reproduce some of the results 
calculated or anticipated earlier in the literature. We also discuss possible applications 
of our findings.

\end{abstract}

\submitto{\jpg}
\maketitle

\date{\today}


\section{Introduction}

   It is well known that the presence of magnetic fields catalyzes the dynamical
breakdown of chiral symmetry even for the weakest attractive interaction
between fermions. This effect has been dubbed as
\emph{magnetic catalysis}~\cite{magcat1,magcat2,magcat3,magcat4}. 
More recently, the dynamical generation of anomalous magnetic moment of the
electron has also been unveiled in~\cite{Ferrer:2009}.
The change in the vacuum structure is governed by the 
the fermion condensate $\langle\bar{\psi} \psi\rangle$, which serves as 
an ideal order parameter to capture the details of this effect. 
At high temperatures, the
structure of the phase transition
becomes even richer. In Ref.~\cite{tempfin1}, 
exploring the magnetic catalysis in massless, weakly coupled QED 
with a magnetic field and temperature, 
within the context of a constant dynamical mass and only considering the lowest Landau level, 
it was established that the chiral symmetry is restored above a critical temperature which is 
much smaller than $\sqrt{eB}$. Under more refined treatment, it was shown in Ref.~\cite{GusShov} that 
symmetry restoration is achieved
for a temperature of the order of the dynamically generated mass of the fermion at zero temperature. The phase transition in this case was found to be of second order~\cite{tempfin2}.
Ref.~\cite{Alexandre} studies the Debye mass
and its dependence as a function of magnetic
field and temperature by means of computing the vacuum polarization tensor in QED.
The work in Ref.~\cite{Sato} deals with the purely thermodynamic part of the one-loop effective 
potential of a fermion in a constant magnetic field. This article is mainly
concerned with the small and large mass expansions for arbitrary values of the magnetic
field and temperature, emphasizing the interpretation of this effective potential
as a dimensionally reduced one from D to D-2 dimensions. Explicit 
expressions are not derived for given hierarchies of temperature or magnetic field strengths. 
This scenario of magnetic fields and finite temperatures is relevant
for astrophysical processes such as supernova explosions as well as early universe 
physics where both the heat bath and
the magnetic field participate as key players in this connection. For example, 
it has been shown by some of us that the 
inclusion of weak magnetic fields enhances the strength of the first order
electroweak phase transition at finite temperature~\cite{Alejandro}. 
Fermion condensation by the effects of a magnetic field could give 
a further boost to these findings. Moreover,  the stability of multi-quark droplets using the NJL model 
at finite temperature and with a homogeneous magnetic field is discussed in Ref.~\cite{Ebert}, showing 
that the later  
promotes the creation of stable droplets and enhances their stability even for quark couplings 
below the value of the bag constant.

On more terrestrial grounds, relativistic heavy-ion collisions at finite impact parameter offer 
an opportunity to test the effects of an external magnetic field in the chiral 
phase transition under extreme conditions~\cite{RHIC}. It has recently been shown that 
in peripheral collisions, a magnetic field of non-negligible strength is 
generated~\cite{McLerran}, the origin of which has two components: the effective angular 
momentum generated by the local imbalance of the momentum carried by the charged 
colliding nucleons~\cite{angmom}, and the addition of currents generated by the 
spectator nucleons which move in opposite directions in the interaction region. 
The effect of a strong magnetic field (in RHIC, magnetic fields can reach intensities 
in the range of $10^2-10^3MeV^2$ for proper times $\tau\lesssim 0.1fm$~\cite{McLerran}) 
on the chiral phase transition has been 
predicted to modify a crossover into a weak first-order transition in the linear sigma model~\cite{lsigma}. 
Nevertheless, since the intensity of this field reduces as $1/\tau^3$~\cite{McLerran}, 
at $\tau\simeq 1fm$, 
the field strength has already reduced two orders of magnitude. Therefore, it is desirable 
to capture the details of the transition over a broad range of values 
of $eB$. This will become even more important in the LHC, where the magnetic fields in 
off-center  heavy-ion  collisions are expected to reach even higher intensities~\cite{kos}.

In this article, we employ the Schwinger's proper-time method to evaluate  
the fermion condensate 
$\langle \bar{\psi} \psi\rangle$ 
at finite temperature in the presence of an external
uniform magnetic field. We
present general results without selecting a hierarchy between
different energy scales involved, namely, temperature $T$, magnetic
field $\sqrt{eB}$ and the bare fermionic mass  $m$.  
We have organized the article as follows. In Sect. II, we calculate
$\langle\bar{\psi} \psi\rangle$ 
at zero temperature in a uniform magnetic field. We also present results
for the limiting cases of $\sqrt{eB} \ll m$ and $\sqrt{eB} \gg m$. 
In Sect. III, the same computation is carried out at finite temperature
without magnetic fields.
After giving a general expression, we
also report analytical expressions in the limit $T \ll m$ as well as 
$T \gg m$. In Sect. IV, we
calculate the fermion condensate in a magnetic field immersed in a
heat bath. Starting from the general results, we consider
special cases of different hierarchies of energy scales involved
($m$, $B$ and $T$), {\em i.e.}, $\sqrt{eB} \gg T \gg m$, $T \gg \sqrt{eB} \gg m$ 
and $T \gg m \gg \sqrt{eB}$. Section V sums up the conclusions.

\section{Condensate in vacuum for $B\neq0$}

The $\langle\bar{\psi} \psi\rangle$-condensate is related to the 
fermion propagator $S(k)$ as
follows~:
\bea
\langle \bar{\psi} \psi \rangle=-Tr \int  \frac{d^4k}{(2\pi)^4} S(k) \;.
\label{chiral1}
\eea
We are interested in calculating the non-dynamical part of this condensate
in the presence of a uniform external magnetic field for a chirally asymmetric 
theory with fermion bare mass $m$. The part of the condensate associated with the
appearance of dynamical masses will be studied elsewhere.

We employ the Schwinger's proper time method~\cite{schwinger} because it is easy 
to be extended to the case of finite temperature in Sect.~IV.
Therefore, assuming the magnetic field to be directed along the third spatial 
direction, we use the expression for the fermion propagator~\cite{schwinger}~:
\bea
S_B(k)&=&-i\int^\infty_0 \frac{d\tau}{\cos(eB\tau)}
             \exp\left[i\tau\left(k^2_{||}+
             \frac{\tan(eB\tau)}{eB\tau} k^2_{\perp}-m^2
             +i\epsilon\right)\right]\nonumber \\
             &\times& \left[\exp({ieB\tau\sigma_3})
              (m+ \gamma\cdot k_{||}\,)-
            \frac{\gamma \cdot k_{\perp}}{\cos(eB\tau)}
             \right]  \;.\nonumber \\
\label{schwingerprop}
\eea
Throughout this work, we use subscript $B$ to denote a quantity in the 
presence of external magnetic field. Moreover, we employ the notation  
$(a\cdot b)_{||}\equiv a^0b^0-a^3b^3$ and 
$(a\cdot b)_{\perp}\equiv a^1b^1+a^2b^2$ for two arbitrary four-vectors 
$a^{\mu}$ and $b^{\mu}$. $\sigma_3$ is the third Pauli matrix.

Following Ref.~\cite{alanchodos}, 
the integration over proper time in Eq.~(\ref{schwingerprop}) can be 
cast in the form a sum which runs over the Landau
level index $l$, namely
\bea
   S_{B}(k)= i \sum^\infty_{l=0} 
           \frac{d_l(\frac{k_\perp^2}{eB})D + 
           d'_l(\frac{k_\perp^2}{eB}) \bar D}{k^2_{||}-2
           l eB-m^2
           + i\epsilon} + \frac{\slsh{k_{\perp}}}{k^2_\perp} \;, 
\label{ferpropsum} 
\eea 
where
$d_l(\alpha)\equiv (-1)^n e^{-\alpha}
L^{-1}_l(2\alpha)$, $d'_n=\partial d_n/\partial \alpha$  \;,
\bea
D &=& (m+\slsh{k_{||}})+ \slsh{k_{\perp}} \frac{m^2-k^2_{||}}{
{k^2_{\perp}}} \;,\qquad
\bar D = \gamma_5 \slsh{u}\slsh{b}(m + \slsh{k_{||}}) \;,
\label{DDe}
\eea
$L_l^m(x)$ are the Associated Laguerre polynomials, and 
$b^\mu$ is a four-vector indicating the direction of the magnetic field.
Vector $u^\mu$ is defined as $(1, \vec{0})$. In case of a heat bath, 
it describes the plasma rest frame.  
Using  the propagator given in Eq.~(\ref{ferpropsum}), the chiral
condensate can be written as 
\bea
\langle\bar{\psi} \psi\rangle_B=
     \frac{-i4m}{(2\pi)^4} \int d^4k \sum_{l=0}^{\infty}
      \frac{(-1)^l e^{-\frac{k_\perp^2}{eB}}L_l^{-1}(2\frac{k_\perp^2}{eB})}
           {k_{||}^2-2leB-m^2+i\epsilon}  \;.
\label{chiral2}
\eea
Starting from this expression, we can establish the equivalence of the
result obtained in~\cite{magcat2}~(a) and~\cite{magcat2}~(d) (Eqs. (9) and~(19), respectively) and~\cite{anguianobashirraya} 
(Eq.~(9)), obtained through distinct routes and written in an entirely 
different fashion. In order to do so, we perform the integral over 
transverse momenta by employing the relation   
\bea
 \int_0^\infty dx e^{-bx}L_n(x)=(b-1)^nb^{-n-1} \;.
\label{chiral2a}
\eea
The expression for the condensate then simplifies to
\bea
 \langle\bar{\psi} \psi\rangle_B   \hspace{-1mm}  =  \hspace{-1mm}    - 
\frac{meB}{2 \pi^2}\sum_{l=0}^\infty (2-\delta_{0l})
 \hspace{-2mm}        \int_0^\infty  \hspace{-4mm} 
       \frac{d k_{3}}
           {\sqrt{k_{3}^2+2leB+m^2}} \;.
\label{chiral2b}
\eea
where we have used  the ``$i\epsilon$'' prescription, that is to say, we
decomposed the integral in the principal and imaginary parts.
Integrating the above expression term by term, we obtain
\bea
\langle\bar{\psi} \psi\rangle_B
    &=&-\frac{2meB}{(2\pi)^2}
    \left\{
    \ln\left(
         \frac{1+\sqrt{1+x_0^2}}{\sqrt{x_0^2}}
        \right)
+2\sum_{l=1}^{\infty}
      \ln\left(
         \frac{1+\sqrt{1+x_l^2}}{\sqrt{x_l^2}}
        \right)
    \right]  \;,
\label{chiral3}
\eea
where $x_l^2=(2leB+m^2)/\Lambda^2$ and $\Lambda$ is the ultraviolet
cutoff. This is exactly the result obtained in
Ref.~\cite{anguianobashirraya} by quantizing directly the solutions to the 
Dirac equation in the external magnetic field. 

We can also show that Eq.~(\ref{chiral2b}) agrees with the result obtained in
Refs.~\cite{magcat2}~(a) and~\cite{magcat2}~(d). 
To see the equivalence, 
notice that when we derive the integral over $k_3$ with respect to $m^2$, the 
remaining integral becomes convergent, acquiring the form
\bea
\int_0^\infty \frac{dk_{3}}{{(k_{3}^2+\mu^2)^{\frac{3}{2}}}}= 
 \int_0^\infty d\eta e^{-\eta \mu^2}\;,\eea
with the appropriate choice of $\mu^2$. On substituting this expression
into Eq.~(\ref{chiral2b}) and performing the sum over 
$l$,~(see Ref.~\cite{landau}) and the integration over $m^2$,  we are led to the expression for the 
condensate given in Refs.~\cite{magcat2}~(a) and~\cite{magcat2}~(d), i.e.,
\bea
\langle\bar{\psi} \psi\rangle_B
    =-\frac{meB}{(2\pi)^2} 
       \int_{\frac{1}{\Lambda^2}}^\infty \frac{d\eta}{\eta}  e^{-\eta m^2}\coth(\eta eB)\;,
\label{chiral3b}
\eea
where $\Lambda$ is the ultraviolet cut-off of
Eq.~(\ref{chiral3}). 
We have thus established the equivalence of Eq.~(\ref{chiral3}) and Eq.~(\ref{chiral3b}), 
In order to calculate purely the magnetic field effect, we need to subtract out the
vacuum piece. Working with Eq.~(\ref{chiral3b}), we obtain  
\bea
\langle\bar{\psi} \psi\rangle_0&=&
       -\frac{m}{(2\pi)^2} 
       \int_{\frac{1}{\Lambda^2}}^\infty \frac{d\eta}{\eta^2}  e^{-\eta m^2}.
\label{chiral4}
\eea
This result is the origin of the quadratic divergence in the condensate in the vacuum 
piece. 
Subtracting out the vacuum part from Eq.~(\ref{chiral3b}), 
the magnetic field dependent fermion condensate is
\bea
\hspace{-0.6cm}  \Delta \left<\overline{\psi} \psi\right>_B&\equiv&
       \left<\overline{\psi} \psi\right>_B
      -\left<\overline{\psi} \psi\right>_0 \nonumber \\
&&  \hspace{-1.5cm} =
\frac{-m}{(2\pi)^2} 
       \int_{0}^\infty 
       \frac{d\eta}{\eta^2} e^{-\eta m^2} 
       \left[eB\eta\coth(eB\eta)-1\right],
\label{chiral6}
\eea
which is finite.

\begin{figure}[t!] 
\vspace{0.4cm}
{\centering
\resizebox*{0.65\textwidth}
{0.30\textheight}{\includegraphics{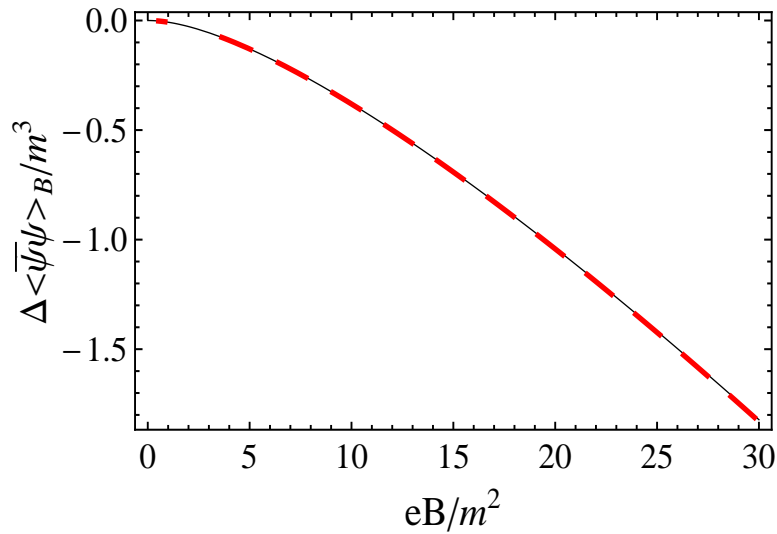}}
\par}
\caption{
Magnetic field dependent fermion condensate in units of $m^3$ 
as a function of $eB/m^2$. \emph{Solid line}: exact result,~Eq.~(\ref{chiral6}). \emph{Thick dashed line}: exact 
result deduced from Eq.~(12) of Ref.~\cite{magcat2}~(a), i.e., Eq.~(\ref{Mir}).
}
\label{fig1pico}
\end{figure}

Note that the above expression  
has been obtained for an arbitrary value of the field strength $eB$, 
as compared to $m$. The numerical plot of the condensate as a function of field strength
is shown in Fig.~\ref{fig1pico} as a solid line. 
In magnitude,  it increases with the
magnetic field strength. The same condensate can be easily extracted 
from Eq.~(15) of the Ref.~\cite{magcat2}~(a). The equivalence can be 
established by taking into account the fact that the condensate 
is given by the derivative of the effective potential, $dV(\rho)/d\rho$, where
$\rho$ is a convenient combination of the fields involved.
The magnetic field dependent condensate would be given by 
\bea
\Delta \left<\overline{\psi} \psi\right>_B = \frac{dV(\rho)}{d\rho} -   
\left[\frac{dV(\rho)}{d\rho}\right]_{B=0} \;.  \label{Mir}
\eea
This expression has been plotted in Fig.~\ref{fig1pico} as a thick dashed line. This
lies exactly on top of the curve representing Eq.~(\ref{chiral6}), 
establishing the correctness of our result.

\begin{figure}[t!] 
\vspace{0.4cm}
{\centering
\resizebox*{0.65\textwidth}
{0.30\textheight}{\includegraphics{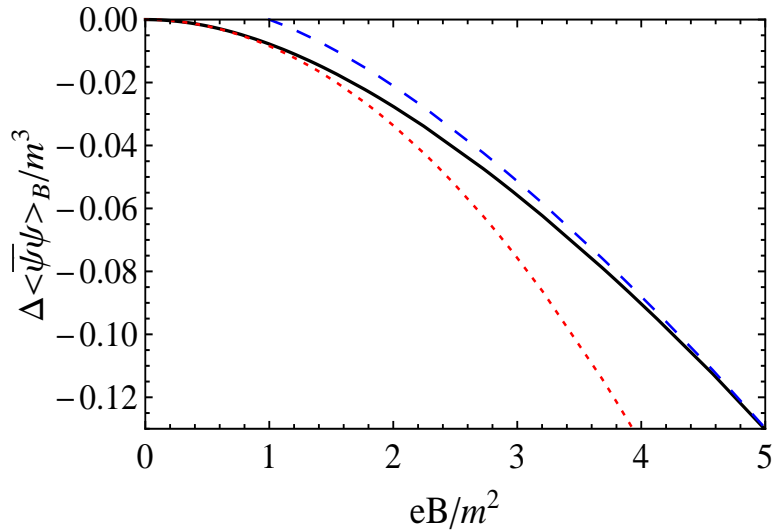}}
\par}
\caption{
Magnetic field dependent fermion condensate in units of $m^3$ 
as a function of $eB/m^2$. \emph{Solid line}: exact result,~Eq.~(\ref{chiral6}). \emph{Dashed line}: strong 
field limit,~Eq.~(\ref{cond31INTENSEeB}). \emph{Dotted line}: weak field 
limit,~Eq.~(\ref{cond31WEAKeB}).
}
\label{fig1}
\end{figure}

From the general expression, we can deduce the
behavior of the condensate for the values of the magnetic field under 
consideration. For example, it is well known that in the strong field limit, 
the main contribution comes from the lowest Landau level (LLL) in  
Eq.~(\ref{chiral3}). In terms of the integral Eq.~(\ref{chiral6}), 
the same  contribution comes from the interval $[1,\infty)$~\cite{landau}. 
Calculating either way, the analytic expression for the result is
\bea
\hspace{-0.5mm}   \Delta \langle\overline{\psi} \psi\rangle_B
   \stackrel{\sqrt{eB} \gg m}{\longrightarrow}
   -\frac{meB}{(2\pi)^2}
    \left[\ln\left(\frac{eB}{m^2}\right)\!-\!\gamma_E\!-1\!\right] \hspace{-0.3mm} .
\label{cond31INTENSEeB}
\eea
This contribution has been displayed by the dashed line in  
Fig.~\ref{fig1}. As expected, it matches on to the exact result in the
intense field limit.

We also obtain a closed expression in the weak field limit~:
\bea
   \Delta \langle\overline{\psi} \psi\rangle_B
   \stackrel{\sqrt{eB} \ll m}{\longrightarrow}-\frac{(eB)^2}{(2\pi)^2m}  \;,
\label{cond31WEAKeB}
\eea
represented by the dotted line in Fig.~\ref{fig1}. It nicely 
reproduces the exact curve for $eB \ll m^2$. Eqs.~(\ref{chiral6}), (\ref{cond31INTENSEeB}), and~(\ref{cond31WEAKeB}) are the main results of this section. 
We now turn our attention to the case of the condensate at finite temperature in 
absence of external magnetic fields.


\section{Condensate at finite temperature and $B=0$}

In this section, we shall study the effect of the heat bath (or thermal fluctuations) at 
temperature $T$ on the fermion condensate in the absence of any external field. On comparison 
with the previous section, we find that the effect of the finite temperature is opposite to 
that of the magnetic field. In order to incorporate the effects of a thermal bath, we use the 
imaginary-time formulation of thermal field theory~(see, for example,~\cite{kapusta}). In this 
formalism, we replace integration over the time component $k_0$ with a sum over discrete 
(Matsubara) frequencies according to the prescription
\bea
   \int\frac{d^4k}{(2\pi)^4}f(k)\rightarrow
   T\sum_n\int\frac{d^3k}{(2\pi)^3}f(\omega_n,{\bf k})
\label{defmatsu}
\eea
where $\omega_n=(2n+1)\pi T$ for fermions, with $n=0,\pm
1,\pm 2,\pm 3\ldots$, and $T$ is the temperature. We shall use the
convenient notation
\bea
   T\sum_n\int\frac{d^3k}{(2\pi)^3}f(\omega_n,{\bf k})
    \equiv \int_\beta f(\omega_n,{\bf k}) \;,
\label{defmatsubeta}
\eea
to denote integration and summation over Matsubara frequencies
$\omega_n$, with $\beta=1/T$.
At finite temperature, the fermion condensate 
is  
\bea
\langle\bar{\psi} \psi\rangle^T&=&
      -4m \int_\beta \frac{1}{\omega_n^2+ \omega_k^2}
=- \frac{m\beta^2}{\pi^2} 
      \int_\beta
      \int_0^\infty ds \ e^{-s\left(\frac{\omega_k}{2\pi T}\right)^2}
      e^{-s\left(n+\frac{1}{2}\right)^2}. 
\label{chiral10}
\eea
The superscript $T$, from now on, denotes a quantity affected by the heat bath. 
Moreover, $\omega_k^2=k^2+m^2$. Note that we have made use of the identity
$ a^{-1} =\int_0^\infty ds \ e^{-s a}$,
which is valid for $Re[a]>0$. As a result, the integral over momenta
is Gaussian and that the sum over Matsubara frequencies gets decoupled from
this integral.   

Since the sum over Matsubara frequencies converges slowly, 
we perform the sum by means of the Poisson resummation formula, 
\bea
   \sum_ne^{-s\left(n+\frac{1}{2}\right)^2} &=&
   \left(\frac{\pi}{s}\right)^{\frac{1}{2}}
   \sum_n(-1)^n e^{-\frac{n^2\pi^2}{s}}.
\label{chiral10a}
\eea
It is well known that once the sum
over Matsubara frequencies is carried out, the result contains the
vacuum contribution. Particularly, in the above expression, the vacuum
contribution  comes  from the term $n=0$. To see this explicitly, 
we only need to replace Eq.~(\ref{chiral10a}) with $n=0$ into
Eq.~(\ref{chiral10}). Thus 
\bea
 \langle\bar{\psi} \psi\rangle^0&=&
     -\frac{m\beta^2}{\pi^{3/2}} 
      \int\frac{d^3k}{(2\pi)^3}
      \int_0^\infty \frac{ds}{s^{\frac{1}{2}}}
      e^{-s\left(\frac{\omega_k}{2\pi T}\right)^2}.
\eea
The integration over $s$ is straightforward, yielding 
\bea
   \langle\bar{\psi} \psi\rangle^0&=&
    -2m \int \frac{d^3k}{(2\pi)^3} \frac{1}{\sqrt{k^2+m^2}},
\label{chiral10aa}
\eea
which corresponds precisely to the fermion condensate in vacuum.

\begin{figure}[t!] 
\vspace{0.4cm}
{\centering
\resizebox*{0.65\textwidth}
{0.30\textheight}{\includegraphics{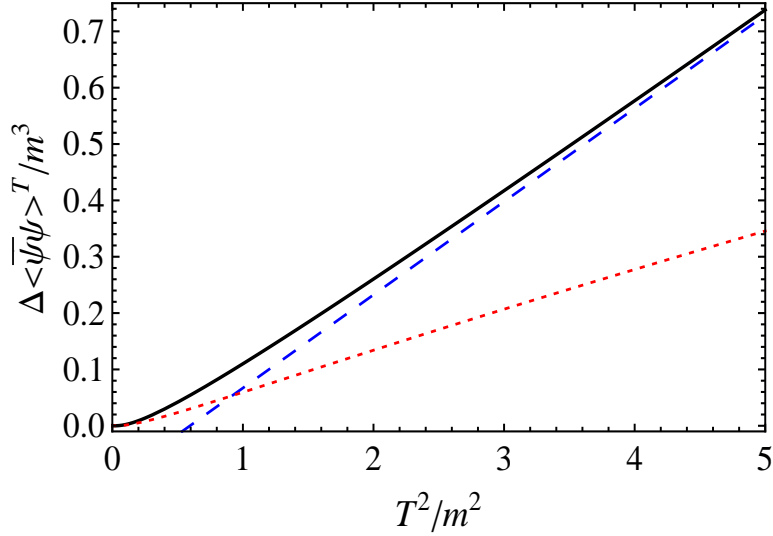}}
\par}
\caption{Temperature dependent fermion condensate in the units of $m^3$ 
as a function of
  $T^2/m^2$. {\it Solid line}: exact result, Eq.~(\ref{chiral10e}). 
   {\it Dashed line}: high temperature limit, Eq.~(\ref{chiral12}). {\it Dotted
     line:} low temperature limit, Eq.~(\ref{lowT31}).} 
\label{fig2}
\end{figure}

Now we proceed to analyze the sum of the remaining terms with $n\neq0$.
These terms contain purely thermal contributions. In order to do so, the
first step is to perform the integral over momentum in
Eq.~(\ref{chiral10}), 
\bea
   \int \frac{d^3k}{(2\pi)^3} e^{-s \frac{k^2}{(2\pi T)^2}}
   = \left( \sqrt{\frac{\pi}{s}} \, T \right)^3  \;.
\label{chiral10b}
\eea
Inserting the above result along with Eq.~(\ref{chiral10a}) 
into  Eq.~(\ref{chiral10}), we find that
\bea
\hspace{-0.5cm}    \Delta\langle\bar{\psi} \psi\rangle^T&=&   
\langle\bar{\psi} \psi\rangle^T - \langle\bar{\psi} \psi\rangle^0  \nonumber \\
&&  \hspace{-1.5cm} =  mT^2\sum_{n\neq0}(-1)^{n+1}\int_0^\infty \frac{ds}{s^2}
   e^{-\left[\left(\frac{m}{2\pi
         T}\right)^2s+\frac{\pi^2n^2}{s}\right]} \;.
\label{chiral10c}
\eea
This result can be further simplified with the help of the identity~\cite{Gradshteyn},
\bea
   \int_0^\infty ds \ s^{-\nu-1}e^{-\gamma s - \frac{\delta}{s}}
   =2\left(\frac{\gamma}{\delta}\right)^{\frac{\nu}{2}}
    K_\nu(2\sqrt{\delta\gamma})\,,
\label{chiral10d}
\eea 
valid for $\mbox{Re}[\delta]>0$ and  $\mbox{Re}[\gamma]>0$, and where $K_\nu(x)$
are the Bessel functions of the second kind. We thus find the temperature dependent 
fermion condensate to be
\bea
\hspace{-0.5cm} \Delta\langle\bar{\psi} \psi\rangle^T&=&
      -
      \frac{ 8mT^2}{(2\pi)^2} 
      \sum_{n=1}^{\infty}\frac{(-1)^n}{n}
      \frac{m}{T}K_1\left( n\frac{m}{T}\right) \;.
\label{chiral10e}
\eea
This is an exact result, valid for all
ranges of temperature and mass. Numerically, it converges rapidly
for increasing values of Matsubara frequencies. It is plotted
as a solid line in Fig.~\ref{fig2}. The condensate is positive and
it rises with increasing temperature.
Notice that the relative sign between the thermal and
magnetic condensates (recall Fig.~\ref{fig1}), indicates that the
heat bath acts opposite to the effect of the magnetic field.

From the exact result, we can extract
the behavior of the condensate
under extreme conditions of high temperature in a 
closed form. To this end, we  make use of the
identity~\cite{htlsum}  
\bea
   \sum_{n=1}^\infty \frac{1}{n}K_1(nz)\cos(n\phi)
   &=&-\frac{z}{4}
   \left[
   \ln\left(\frac{z}{4\pi}\right)+\gamma_E-\frac{1}{2}
   \right]
+\frac{1}{z}
   \left[
   \frac{1}{4}\phi^2-\frac{\pi}{2}\phi+\frac{\pi^2}{6} 
   \right]
   \nonumber\\
   &&\hspace{-15mm}-\frac{\pi}{2z}\sum_{l\neq0}
   \left[
    \sqrt{z^2+(\phi-2\pi l)^2}
-\vert\phi-2\pi l\vert
   -\frac{z^2}{4\pi \vert l\vert}
   \right].
\label{chiral11a}
\eea
Notice that when $\phi=\pi$  the sum on the {\em l.h.s} reduces to the sum in 
Eq.~(\ref{chiral10e}). In
the high temperature limit, {\it i.e.,} $m/T \ll 1$ (or $z\ll1$) the
first and third terms on the {\em r.h.s} of the identity  vanish. 
This allows us to evaluate the fermion condensate analytically in this  limit~:
\bea
\Delta\langle\bar{\psi} \psi\rangle^T    &\stackrel{T\gg m}{\longrightarrow}& 
  \frac{ mT^2}{6}  \;. \label{chiral12}
\eea
It is displayed in Fig.~\ref{fig2} with a dashed line. The exact result
aligns itself with  Eq.~(\ref{chiral12}) in the high temperature limit.
The low temperature behavior for the temperature dependent condensate can be
obtained  from  Eq.~(\ref{chiral10e}) by means of an asymptotic
expansion for the Bessel functions near  infinity. We can then
perform the sum over $n$ term by term. We again obtain a closed expression
in this regime,    
\bea
   \Delta \left<\overline{\psi} \psi\right>_T
     &\stackrel{T\ll m}{\longrightarrow}& 
     \frac{1}{2}\left(\frac{2mT}{\pi}\right)^{\frac{3}{2}}e^{-\frac{m}{T}}  \;.
\label{lowT31}
\eea
This result is depicted in Fig.~\ref{fig2} by the dotted line. Note that
the high temperature limit,  Eq.~(\ref{chiral12}),
breaks down at lower values of temperature and is taken over by Eq.~(\ref{lowT31})
to mimic the exact result.
Eqs.~(\ref{chiral10e}),~(\ref{chiral12}), and~(\ref{lowT31}) constitute the main results 
of this section. 
The combined effect of an external uniform magnetic field and a heat bath on the 
fermion condensate is studied in the next section.

\section{Condensate in a heat bath with $B\neq0$}

In previous sections, we have observed that the effects of the external uniform magnetic field 
and the thermal bath on the fermion condensate are diametrically opposed. A natural and more 
relevant scenario is to study the combined effect of these two antagonic agents on 
$\langle\bar{\psi} \psi\rangle$. We apply the procedure similar to the one developed in the 
previous two sections and arrive at the following expression~: 
\bea
\hspace{-0.5cm}   \Delta \langle\bar{\psi} \psi \rangle_B^T&=&
   -\frac{meB}{\pi^2}
    \sum_{l=0}^\infty(2-\delta_{0l})  
\sum_{n=1}^\infty (-1)^n K_0\left(n \frac{\sqrt{2leB+m^2}}{T}\right)\nonumber\\
& +&
\Delta \left<\overline{\psi} \psi\right>_B \;.
\label{chiral13}
\eea
Notice that the first
term takes into account the effects of thermal bath and the external
magnetic field on the fermion condensate, while the second corresponds
to the condensate in vacuum with a background magnetic field. This is one
of the main results of our paper.

\begin{figure}[t!] 
\vspace{0.4cm}
{\centering
\resizebox*{0.65\textwidth}
{0.25\textheight}{\includegraphics{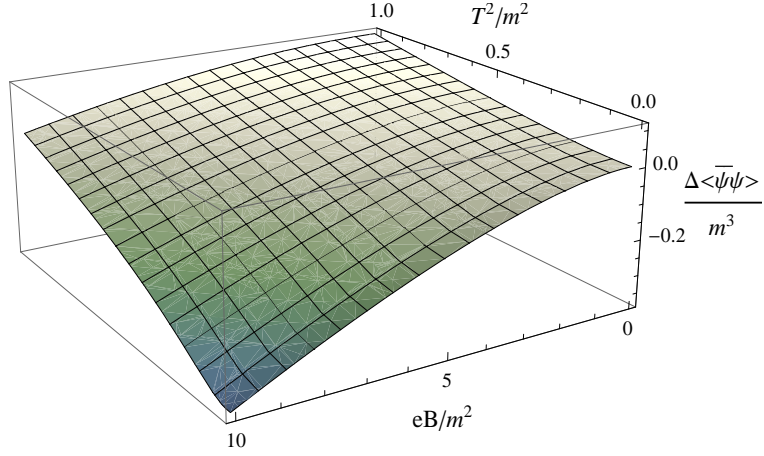}}
\par}
\caption{Temperature and magnetic field dependent fermion condensate as a function of
  $eB/m^2$ and $T^2/m^2$. } 
\label{fig3}
\end{figure}

\begin{figure}[t!] 
\vspace{0.4cm}
{\centering
\resizebox*{0.6\textwidth}
{0.45\textheight}{\includegraphics{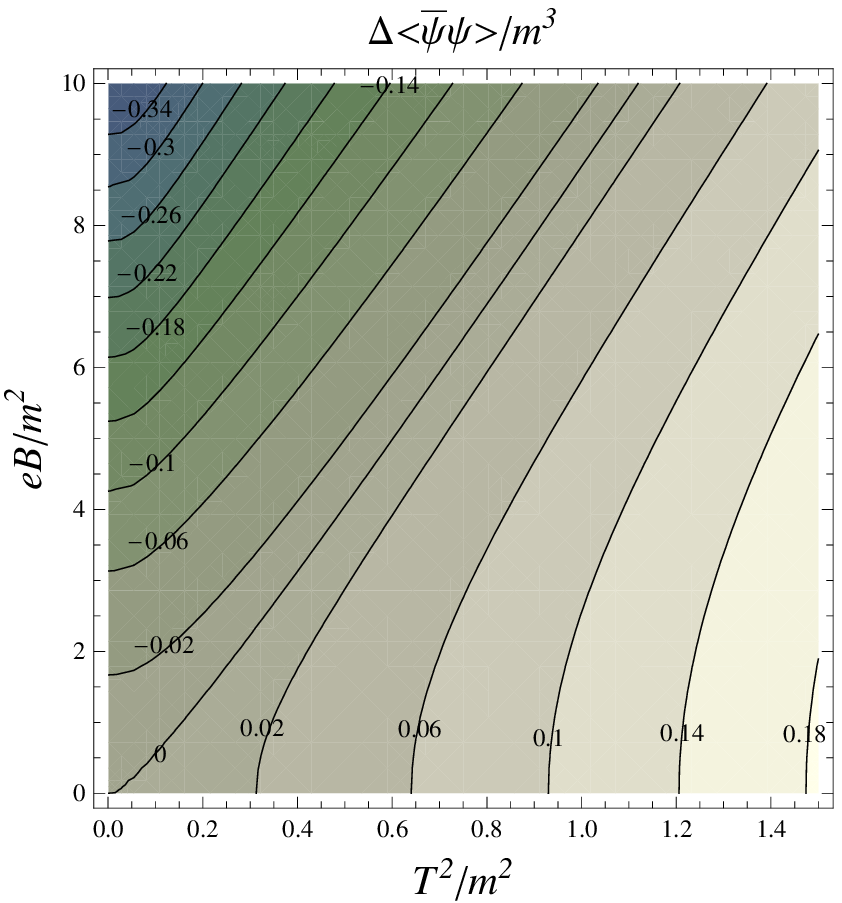}}
\par}
\caption{Contour plot for the temperature and magnetic field dependent fermion condensate as a function of
$eB/m^2$ and $T^2/m^2$.} 
\label{fig4}
\end{figure}

Eq.~(\ref{chiral13}) can also be cast in the
following form~: 
\bea 
  \Delta  \left<\overline{\psi} \psi\right>_B^T&=&
   \frac{2meB}{(2\pi)^2}\sum_{n=1}^\infty(-1)^{n+1}\int_0^\infty\frac{ds}{s}
   e^{-\left[\frac{m^2}{(2\pi T)^2}s+\frac{\pi^2n^2}{s}\right]} 
   \coth\left(\frac{eBs}{(2\pi T)^2}\right)
   \nonumber \\ 
    &+&  \Delta \left<\overline{\psi} \psi\right>_B   \;.
\label{chiral13bb}
\eea
This exact expression can also be written in terms of the Jacobi $\Theta_4$ function
originating from the sum over $n$,~\cite{Gusynin-Incera-Ferrer:2003}
as follows~:
\bea
     \Delta  \left<\overline{\psi} \psi\right>_B^T&=&
   -\frac{meB}{(2\pi)^2}\int_0^\infty\frac{ds}{s}\left[\Theta_4(0;e^{-\frac{\pi^2}{s}}) -1\right]
   e^{-\frac{m^2}{(2\pi T)^2}s} 
   \coth\left(\frac{eBs}{(2\pi T)^2}\right)
   \nonumber \\ 
    &+&  \Delta \left<\overline{\psi} \psi\right>_B   \;.
\label{chiral13bb-theta}
\eea
The exact form is plotted in Fig.~\ref{fig3}. Note that at
high temperatures and weak magnetic field (the farthest edge of the cube), the
condensate is positive. For low temperatures and intense magnetic fields
(the closest edge of the cube), playing the dominant and opposite role, the magnetic 
field pulls down the condensate to its large negative values. In the region where
the condensate does not deviate much from its zero value (from the right edge of the
cube towards the left edge), temperature and magnetic fields manage to nullify 
the effect of each other. This effect is most visible in the contour plot displayed
in Fig.~\ref{fig4}.

Starting from Eq.~(\ref{chiral13bb}), we can analyze different scenarios of relative 
strengths of the mass $m$, the magnetic field $eB$ and the temperature $T$. 
In the following sub-sections, we discuss various hierarchies of interest,
of the energy scales involved~: 
intense magnetic fields, {\em i.e.}, {$  m \ll T \ll \sqrt{eB}$}, intermediate magnetic
fields, namely {$  m \ll \sqrt{eB} \ll T$},  
and  weak magnetic fields, {$ \sqrt{eB} \ll m \ll T$}.

\subsection{Strong Field}

\begin{figure}[t!] 
\vspace{0.4cm}
{\centering
\resizebox*{0.65\textwidth}
{0.30\textheight}{\includegraphics{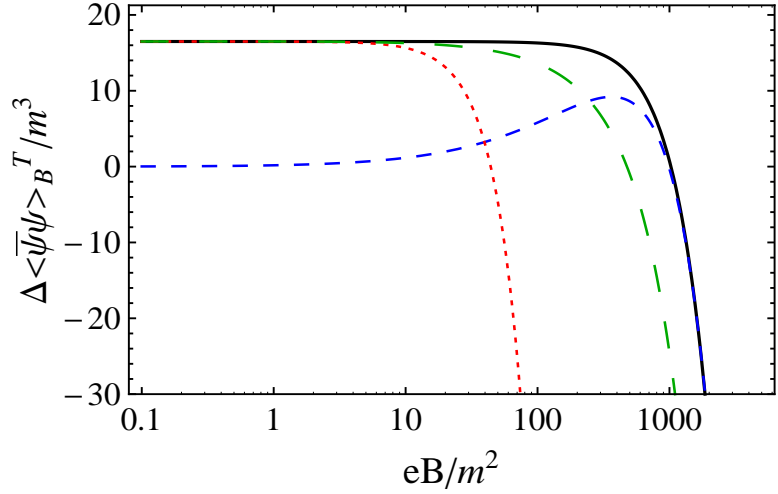}}
\par}
\caption{Temperature and magnetic field dependent fermion condensate as a
  function of $eB/m^2$ at fixed $T$. {\it Solid line}: exact (numerical) result, Eq.~(\ref{chiral13bb}). 
  {\it Dashed line}: strong field limit, Eq.~(\ref{chiral13c}).
   {\it Long Dashed  line}: intermediate field limit (numerical), Eq.~(\ref{chiral13ee}).
   {\it Dotted line}: weak field limit, Eq.~(\ref{capodebi5}).}
\label{fig5}
\end{figure}

Let us begin by considering the following hierarchy among the energy scales involved:
$ m \ll T \ll \sqrt{eB} $. 
In such a scenario,  the main
contribution comes from the LLL, i.e.,
$l=0$ in Eq.~(\ref{chiral13}). In this regime, the condensate is given by    
\bea
 \Delta  \langle \bar{\psi} \psi \rangle_B^T &=&
   -\frac{meB}{\pi^2}\sum_{n=1}^\infty (-1)^n
       K_0\left(n \frac{m}{T}\right) 
-\frac{meB}{(2\pi)^2}
    \left[\ln\left(\frac{eB}{m^2}\right)-\gamma_E-1\right] \;.
\label{chiral13a}
\eea
Furthermore, we again use Eq.~(\ref{chiral11a})
to perform the high temperature expansion  ($m \ll T$)~:
\bea
   2\sum_{n=1}^\infty (-1)^n K_0\left(n\frac{m}{T}\right)
  &\stackrel{m\ll T}{\rightarrow}&
  \gamma_E-1+\ln\left(\frac{m}{\pi T}\right)\;.
\label{chiral13b}
\eea
Substituting this expression into Eq.~(\ref{chiral13a}) and using
Eq.~(\ref{cond31WEAKeB}) yields the fermion condensate 
\bea
 \Delta    \left<\overline{\psi} \psi\right>_B^T=
   -\frac{meB}{(2\pi)^2}
    \left[\ln\left(\frac{eB}{\pi^2 T^2}\right)+ \gamma_E-3\right]  \;.
\label{chiral13c}
\eea
This is drawn as the dashed line in Fig.~\ref{fig5}
against the exact result for a given value of temperature, represented
by the solid line. Note that it fits the exact result very well when
the magnetic field intensity is sufficiently  large as compared to
other energy scales in the problem. However, this LLL approximation 
breaks down
when magnetic fields  are lowered, and we have to resort to other
approximation schemes as detailed in the sub-sections to follow.

\subsection{Field of Intermediate Intensity}

In this regime, the temperature  dominates over any other energy scale.
Therefore, we expect that the main contribution to fermion
condensate comes from the thermal bath. In order to confirm this, we 
take the high temperature limit of Eq.~(\ref{chiral13}). Employing
Eq.~(\ref{chiral13b}), we arrive at
\bea
   \Delta\left<\overline{\psi} \psi\right>_B^T&=&
   -\frac{2meB}{(2\pi)^2}\sum_{l=0}^\infty (2-\delta_{l0})
\left[
   \ln\left(\frac{\sqrt{2leB+m^2}}{4\pi T}\right)
   +\gamma_E-1
   \right]\nonumber \\
   &-& \frac{meB}{(2\pi)^2}
    \left[\ln\left(\frac{eB}{m^2}\right)-\gamma_E-1\right] \;,
\label{chiral13e}
\eea
where we have also made use of the fact that $\sqrt{e B} \gg m$,
Eq.~(\ref{cond31INTENSEeB})
As the magnetic field provides an intermediate energy scale, 
fermions can access all the Landau
levels and we have to perform the sum over all of them.
This is achieved by invoking the procedure outlined in
Ref.~\cite{landau}~:
\bea
   \Delta\left<\overline{\psi} \psi\right>_B^T&=&
   \frac{meB}{(2\pi)^2}\int_{\frac{3}{2\pi^2T^2}}^\infty
   \frac{d \eta}{\eta}e^{-\eta m^2}\coth(\eta eB)
    \nonumber \\
   &-& \frac{meB}{(2\pi)^2}
    \left[\ln\left(\frac{eB}{m^2}\right)-\gamma_E-1\right]  \;.
\label{chiral13ee}
\eea
Notice that we have introduced an infrared cutoff $(2/3) \pi^2 T^2$. This
is due to the fact that in the limit $eB\rightarrow 0$, each component of 
the transverse momentum contributes to the thermal bath with 
a factor of $(1/3)\pi^2 T^2$. 
In this kinematical region, we cannot perform any other approximation 
and hence a closed expression alludes us.  
In Fig.~\ref{fig5},
we plot the numerical results for Eq.~(\ref{chiral13ee})  in the form of the
long dashed line. It nicely captures the correct intermediate field range.

\subsection{Weak field}

We finally consider the weak field scenario $\sqrt{eB} \ll m \ll T$.
In this case,  we have to perform sum over all Landau levels, since the gap 
between energy levels is small and thermal fluctuations can bring fermions 
from one level to a higher one easily. Notice that
Eq.~(\ref{chiral11a}) no longer provides a useful starting point because 
the sum over Landau levels exhibits a divergence. Nevertheless,
we can  perform a weak field expansion in the fermion propagator, 
Eq.~(\ref{ferpropsum}), 
which allows to perform the summation over Landau levels. 
The resulting fermion propagator is then written as a power series in $eB$. 
Up to order $(eB)^2$ it reads~\cite{taiwaneses} 
\bea
    S_B(k)   &\approx&  \frac{\slsh{k}+m}{k^2-m^2}
      +\frac{\gamma_5\slsh{u}\slsh{b}(\slsh{k}_{||}+m)eB}{(k^2-m^2)^2}
      \nonumber \\      &-& 
\frac{2(eB)^2k_\perp^2}{(k^2-m^2)^4}
       \left(m+\slsh{k}_{||}+\slsh{k}_\perp\frac{m^2-k^2}{k_\perp^2}\right).
\label{capodebi1}
\eea 
Substituting this expansion into Eq.~(\ref{chiral1}), the
fermion condensate in the weak field limit is given by 
\bea
\langle\bar{\psi} \psi\rangle_B&=&
     -i Tr  \int \frac{d^4k}{(2\pi)^4} 
     \left[
      \frac{\slsh{k}+m}{k^2-m^2}
     +\frac{\gamma_5\slsh{u}\slsh{b}(\slsh{k}_{||}+m)eB}{(k^2-m^2)^2}\right.\nonumber\\
     &&-\frac{2(eB)^2k_\perp^2}{(k^2-m^2)^4}
\left.
       \left(m+\slsh{k}_{||}+\slsh{k}_\perp\frac{m^2-k^2}{k_\perp^2}\right)    
     \right] \;.
\label{capodebi2}
\eea
Notice that the term linear in the magnetic field vanishes
when we take the trace. Thus,
\bea
\hspace{-0.5cm} \langle\bar{\psi} \psi\rangle_B&=&
     -i 4m  \int \frac{d^4k}{(2\pi)^4} 
     \left[
      \frac{1}{k^2-m^2}
    - \frac{2(eB)^2k_\perp^2}{(k^2-m^2)^4}    
     \right].
\label{capodebi3}
\eea
With the prescriptions of the imaginary time formalism, Eq.~(\ref{defmatsu}),  the fermion
condensate in the weak field limit at finite temperature can 
be written as   
\bea
\langle\bar{\psi} \psi\rangle^T_B&=&
     -4m  T\sum_n\int \frac{d^3k}{(2\pi)^3} 
     \left[
      \frac{1}{\omega_n^2+k^2+m^2}
+\frac{2(eB)^2k_\perp^2}{(\omega_n^2+k^2+m^2)^4}    
     \right]\;.
\label{capodebi4}
\eea
Notice that the first term was calculated in
Eq.~(\ref{chiral10}) with the result shown in
Eq.~(\ref{chiral12}). Given the hierarchy of energy scales in this limit, the
thermal component that comes from the second term is negligible. 
Therefore, the main contribution comes from the vacuum part. 
After some routine algebra, the condensate in this 
limit is given by the following closed expression~:
\bea
\hspace{-1cm}  \Delta\left<\overline{\psi} \psi\right>^T_B&=&
    \frac{mT^2}{6}
      \left[
      1-\frac{7\zeta(3)}{8\pi^4}\frac{(eB)^2}{T^4} 
-\frac{1}{(2\pi^2)}\frac{(eB)^2}{m^2T^2}
      \right]  \;, 
\label{capodebi5}
\eea
where $\zeta(x)$ is the Riemann zeta function.
This expression is depicted by the dotted line in Fig.~\ref{fig5}, which
lies on top of the exact result in the limit of weak magnetic fields.
With increasing values of the magnetic field, this result starts becoming 
more and more inaccurate while the intermediate field limit,~Eq.~(\ref{chiral13ee}), 
still preserves its validity. For intense magnetic fields,~Eq.~(\ref{chiral13c}) takes
over as the adequate description of reality.

\section{Conclusions}

 The simultaneous presence of finite temperature and external magnetic fields is
a realistic situation in the physics of the early universe~\cite{Alejandro}, astronomical objects
and even heavy ion collisions~\cite{RHIC}. Therefore, it is important to estimate the effect
of these conditions on the fermion mass and the fermion-anti-fermion condensate.
Moreover, in all the practical situations 
in cosmology, astrophysics, or heavy ion collisions, charged fermions have
a bare mass as long as the electroweak phase transition has taken place.
Its contribution to the mass gap or the  $\langle\bar{\psi} \psi\rangle$ condensate is 
an important question whose relevance has been highlighted in several recent
works, both in the vacuum,~\cite{vacuum-bare}, as well as in the presence of an external magnetic
field,~\cite{field-bare}.

In this article, we present a detailed study of this effect in QED on the $\langle\bar{\psi} \psi\rangle$ condensate
for the bare fermionic mass $m$ in the absence of self interactions. Most importantly,
our results are valid for an arbitrary hierarchy of the energy scales involved, namely, the 
fermion mass $m$, the external magnetic field strength $\sqrt{eB},$ and the temperature $T$. This is achieved by carrying out the sum over 
all the Landau levels and all the Matsubara frequencies. To the best of our knowledge, no such 
complete calculation exists so far in the literature. The effect of the temperature and the
magnetic fields are diametrically opposed, thus tending to nullify each other.
We also take several physically 
relevant limits of our general results and arrive at closed
expressions for particular regimes of $T$, $\sqrt{eB}$ and $m$. This analysis explicitly
reveals the domain of validity of each and every approximation employed~:
$\sqrt{eB} \gg T \gg m$, $T \gg \sqrt{eB} \gg m$ and $T \gg m \gg \sqrt{eB}$. We also present
exact results for the cases when either temperature or the magnetic field is absent.
For special cases, we reproduce the results already known in the literature.  The main 
results of our article are~: Eqs.~(\ref{chiral6}), (\ref{cond31INTENSEeB}), and~(\ref{cond31WEAKeB})
for $T=0,~B\neq0$, Eqs.~(\ref{chiral10e}), (\ref{chiral12}), and(~\ref{lowT31}) for $T\neq0,~B=0$
and Eqs.~(\ref{chiral13bb}), (\ref{chiral13c}), (\ref{chiral13ee}), and~(\ref{capodebi5}) for
$T\neq0,~B\neq0$. A natural next step is to include the dynamical interaction effects
while still preserving the arbitrariness of temperature as well as external magnetic field.
All this is for future.

\ack

We are grateful for helpful comments of E. Ferrer, V.P. Gusynin and V. de la Incera
on the draft version of this article.
A.A. whishes to thank the kind hospitality of both faculty and staff in
IFM-UMSNH during a sabbatical visit and the financial support of DGAPA-UNAM
under PAPIIT grant No. IN116008. A.B. and A.R. acknowledge COECyT, CIC an CONACyT grants
while A.S. acknowledges postdoctoral CONACyT fellowship.



\section*{References}
\begin{thebibliography}{55}

\bibitem{magcat1} (a) K G Klimenko 1992 {\it Z. Phys.} {\bf C54} 323; 

(b) K G Klimenko 1992 {\it Theor. Math. Phys.} {\bf 89} 1161  [english translation of the russian
paper K G Klimenko 1991 {\it Teor. Mat. Fiz.} {\bf 89} 211]; 

(c) K G Klimenko 1992 {\it Theor. Math. Phys.} {\bf 90} 1  [english translation of the russian paper
K G Klimenko 1992 {\it Teor. Mat. Fiz.} {\bf 90} 3].

\bibitem{magcat2} (a) V P Gusynin, V A Miransky and I A Shovkovy 1995
{\it Phys. Lett. B} {\bf 349} 477; 

(b) V P Gusynin, V A Miransky and I A Shovkovy 1995 {\it Phys. Rev. D} {\bf 52} 4718; 

(c) V P Gusynin, V A Miransky and I A Shovkovy 1995 {\it Phys. Rev. D} {\bf 52} 4747; 

(d) V P Gusynin, V A Miransky and I A Shovkovy 1996 {\it Nucl. Phys. B} {\bf 462} 249; 

(e) V P Gusynin, V A Miransky and I A Shovkovy 1999 {\it Nucl. Phys. B} {\bf 563} 361; 

(f) V P Gusynin, V A Miransky and I A Shovkovy 1994 {\it Phys. Rev. Lett.} {\bf 73} 3499.

\bibitem{magcat3} (a) D-S Lee, C N Leung and Y J Ng 1997 {\it Phys. Rev. D} {\bf 55} 6504; 

(b) D K Hong 1998 {\it Phys. Rev. D} {\bf 57} 3759; 

(c) E J Ferrer and V de la Incera 2000 {\it Phys. Lett. B} {\bf 481} 287.


\bibitem{magcat4} (a) A Ayala, A Bashir, A Raya and E. Rojas 2006 {\it Phys. Rev. D}
{\bf 73} 105009; 

(b) A Ayala, A Bashir, A Raya and E. Rojas 2008 {\it Phys. Rev. D} {\bf 77}  093004; 

(c) N Sadooghi and  K Sohrabi Anaraki 2008  {\it Phys. Rev. D} {\bf 78} 125019.

\bibitem{Ferrer:2009} E J  Ferrer and V  de la Incera 2009 {\it Phys. Rev. Lett.} {\bf 102} 050402;

E J  Ferrer and V  de la Incera 2009 {\it Nucl. Phys. B} {\bf 824} 217.

\bibitem{tempfin1} D S  Lee, C N Leung and Y J Ng 1997 {\it Phys. Rev. D} {\bf 55} 6504. 

\bibitem{GusShov} V P Gusynin and I A Shovkovy 1997 {\it Phys. Rev. D} {\bf 56} 5251.

\bibitem{tempfin2} D S  Lee, C N Leung and Y J Ng 1998 {\it Phys. Rev. D} {\bf 57} 5224. 

\bibitem{Alexandre} J Alexandre 2001 {\it Phys. Rev. D} {\bf 63} 073010.

\bibitem{Sato} H-T Sato 1998 {\it J. Math. Phys.} {\bf 39} 4540.

\bibitem{Alejandro}
A S\'anchez, A Ayala and G Piccinelli 2007 {\it Phys. Rev. D} {\bf 75} 043004.

\bibitem{Ebert} D Ebert and K G Klimenko 2003 {\it Nucl. Phys. A} {\bf 728} 203;

K G Klimenko and D Ebert, 2005 {\it Phys. At. Nucl.} {\bf 68} 124. 

\bibitem{RHIC} For a recent and comprehensive review, see the proceedings of the workshop on high pt physics at LHC PoS(LHC07).

\bibitem{McLerran} D E Kharzeev, L D McLerran and H J Warringa 2008
{\it Nucl. Phys. A} {\bf 803} 227.

\bibitem{angmom} (a) Z-T Liang and X-N Wang 2005 {\it Phys. Rev. Lett.} {\bf 94} 102301;

(b) F Becattini, F Piccinini and J Rizzo 2008 {\it Phys. Rev. C} {\bf 77} 024906.

\bibitem{lsigma} E S Fraga and A J Mizher 2008 {\it Phys. Rev. D} {\bf 78} 025016.

\bibitem{kos}  V Skokov, A Illarionov and V Toneev \emph{Estimate of the magnetic field strength in heavy-ion collisions} e-pront: arXiv:0907.1396 [nucl-th].

\bibitem{schwinger}
J Schwinger 1951 {\it Phys. Rev.} {\bf 82} 664.


\bibitem{alanchodos}
A Chodos, K Everding, D A Owen 1990 {\it Phys. Rev. D} {\bf 42}, 2881. 

\bibitem{anguianobashirraya}
M de J Anguiano-Galicia, A Bashir and A Raya 2007 {\it Phys. Rev. D} {\bf 76} 127702.


\bibitem{landau}
V B Berestetskii, E M Lifshitz and L P Pitaevskii 1982 {\it Quantum
  Electrodynamics} (Pergamon Press).






\bibitem{kapusta} J I Kapusta 1989 \emph{Finite Temperature Field Theory} (1st. edition, Cambridge University Press). 

\bibitem{Gradshteyn} I S Gradshteyn and  I M  Ryzhik 2000 {\em Table of Integrals, Series and Products}  (Academic Press) p. 567.


\bibitem{htlsum}
P N  Meisinger and M C  Ogilvie 2002 {\it Phys. Rev. D} {\bf 65} 056013. 


\bibitem{Gusynin-Incera-Ferrer:2003} E J  Ferrer, V P  Gusynin and V  de la Incera 2003
{\it Eur. Phys. J. B} {\bf 33}, 397.

\bibitem{taiwaneses}
T-K Chyi, C-W Hwang, W F  Kao, G-L Lin, K-W  Ng and J-J. Tseng 2000
{\it Phys. Rev. D} {\bf 62}, 105014.



\bibitem{vacuum-bare}  L Chang, Y-X Liu, M S Bhagwat, C D Roberts and S V Wright 2007
{\it Phys. Rev. C} {\bf 75}, 015201.

\bibitem{field-bare} S-Y Wang 2008 {\it Phys. Rev. D} {\bf 77}, 025031; 

K G Klimenko and V Ch Zhukovsky 2008 {\it Phys. Lett. B} {\bf 665} 352;

E Rojas \emph{Radiative and non-perturbative corrections to the electron mass and the anomalous magnetic 
moment in the presence of an external magnetic field of arbitrary strength.}
e-Print: arXiv:0811.1066 [hep-ph].


\end{thebibliography}
\end{document}